\documentclass{llncs}
\usepackage{makeidx}  % allows for indexgeneration
\usepackage{amssymb} % for drawing double-barred R, Z, C, etc.
\usepackage{epsfig}

\begin{document}
\title{Equilibrium Point in Quantum Physics\\and its Inspiration to Game Theory}
\author{Xiaofei Huang}
\authorrunning{Xiaofei Huang}   % abbreviated author list (for running head)
\tocauthor{Xiaofei Huang}
\institute{eGain Communications, Mountain View, CA 94043, USA\\
\email{jhuang@egain.com}}

\maketitle              % typeset the title of the contribution

\begin{abstract}
Most of atoms and molecule found in nature are capable of evolving towards and staying at their ground states, 
	the lowest energy states.
This paper offers a global optimization approach to understand the ground state as the equilibrium point of an $n$-player game under cooperation.
With the same approach, Nash equilibrium can be viewed as the equilibrium point under competition.
The former can offer higher payoffs and stability of game playing than the later.
It is truly an inspiration from nature for us to build societies for quality and stability under cooperation rather than competition.
\end{abstract}

\section{Introduction}
The quantum world has many bizarre behaviors~\cite{Tegmark01,Seife05}, such as wave-particle duality and Heisenberg uncertainty principle.
One of them is why most of atoms and molecule found in nature are capable of evolving towards and staying at their ground states, 
	the lowest energy states.
Otherwise, no atom and no molecule will be stable in nature and the world would be completely out of order.

The ground states, and the stationary states in general, are described by Schr\"{o}dinger equation in quantum mechanics~\cite{Messiah99}.
However nobody knows any deeper principle underlying the equation.
This paper attempts to show that
	the equation can be derived from a global optimization algorithm, called cooperative optimization~\cite{HuangBookCCO}.
Indeed, from a computational point of view, 
	each atomic system needs to follow a global optimization process 
	so that it can find its ground state, the lowest energy state. 

Cooperative optimization describes a system with multiple agents, each with its own objective, 
	working together under cooperation to make a joint decision to optimize the collected objectives of individuals.
The same language of mathematics can be used to describe a molecule of multiple atoms and a game of multiple players. 
Cooperative optimization offers a mathematical framework 
	to understand the computational properties of multi-agent systems under different competition and cooperation strategies.

Specifically, if every agent in a multi-agent system accepts only the best action and rejects all sub-optimal ones for the agent,
	it leads to a direct competition among the agents. % because their objectives are commonly not aligned with each other.
If an equilibrium under this kind of competition is reached,
	it will be shown in Section~3 that it is a Nash equilibrium~\cite{nash50} defined in game theory.
Nash equilibrium is a key concept in game theory to understand games with rational players.

In contrary to the above scenario, 
	if every agent is less aggressive at competition and is willing to accept sub-optimal actions to some degree,
	it leads to a cooperation among the agents.
If an equilibrium under this kind of cooperation is reached,
	it will be shown in Section~2 that it is a stationary state described by Schr\"{o}dinger equation in quantum mechanics.
Compared with competition, cooperation can greatly increase the possibility for a multi-agent system to find its global optimal equilibrium state.	

Taking a molecule with multiple atoms as an example,  
	if every atom is extremely aggressive at fighting for the best location in terms of potential energy described by classical physics,
	the molecule follows a local optimization process and will get stuck into one local minimum energy state or another, 
	rather than the global one.
Therefore, the molecule is not stable because it may have an enormous number of local minimal energy states and its final state can be any one of them, 
	sensitive to its initial configuration and perturbations during the process.
	
Contrary to that, if every atom accepts sub-optimal locations 
	to a certain degree while keeping the best location of the highest acceptance,
	the molecule tends to follow a global optimization process and evolves towards the lowest energy state (the global minimal energy state).
In this case, the molecule becomes stable, in-sensitive to its initial configuration and perturbations during the process.
At the end, all the atoms in the molecule jointly get the best possible locations both in a global sense and in an average sense.

If each atom accepts only the best location, 
	it will have a precise location in space at any given time instance.
That is the picture of classic physics at understanding the world.
Different from that, if each atom accepts all locations, the best and sub-optimal ones, proportional to their goodness,
	its location will be spread in space with a probability-like distribution.
That is exactly the physical reality of the quantum world.

Accepting sub-optimal actions as a generic decision-optimization strategy is truly an inspiration of the quantum world.
It offers the definition of another type of equilibrium points of $n$-player games besides Nash equilibrium.
The quantum world suggests us that,
	at a Nash equilibrium, if every player gives away some payoff 
	by accepting sub-optimal actions to some degree, 
	each of them may actually receive a better return at a new equilibrium point than the original one.
At the same time, the game playing may become more stable since the number of its equilibrium points can be reduced remarkably.
Sometimes, it may be reduced to a single one corresponding to the social optimum.

\section{A globalization approach to quantum mechanics}
%%%%%%%%%%%%%%%%%%%%%%%%%%%%%%%%%%%%%%%%%%%%%%%%%%%%%%%%%%%%%%%%%%%

%The global optimization algorithm is defined by the cooperative optimization theory~\cite{HuangBookCCO}.
%The theory is for a mathematical understanding of ubiquitous cooperative behaviors in nature and translating them into optimization algorithms.
Given a society with $n$ individuals, assume that 
	the objective of individual $i$ is described as minimizing a function $E_i(x)$.
A simple form of cooperative optimization is defined as an iterative computation 
	of each individual's expected returns described by a function $\Psi_i (x_i, t)$, for $i=1,2,\ldots,n$, as follows:
\begin{equation}
\Psi_i (x_i, t) = \sum_{\sim x_i} \left( e^{-E_i(x)/\hbar} \prod_{j \not= i} p_j(x_j, t-1) \right)  \ , 
\label{cooperative_optimization_general3}
\end{equation}
where $\sum_{\sim x_i}$ stands for the summation over all variables except $x_i$.
$\hbar$ is a constant of a small positive value.
$p_i(x_i, t)$ is a probability-like function for picking action $x_i$ proportional to $\left(\Psi_i(x_i,t)\right)^{\alpha} (\alpha > 0)$, i.e.,
\begin{equation}
p_i(x_i, t) = \left(\Psi_i(x_i,t)\right)^{\alpha} / Z_i(t) \ , 
\label{computingP}
\end{equation}
 where $Z_i(t)$ is a normalization factor defined as $Z_i(t) = \sum_{x_i} \left(\Psi_i(x_i,t)\right)^{\alpha}$.

The larger the parameter $\alpha$ is, the more aggressive each individual is at minimizing his own objective function $E_i(x)$.
At the same time, the game tends to have more equilibrium points.
However, the chance for the society to reach the social (global) optimum
	is only peaked at a certain value of $\alpha$, neither too large nor too small.
In this case, each individual in the game compromises his best action by accepting sub-optimal actions to some degree,
	different from the case when $\alpha \rightarrow \infty$ where only the best action is accepted (see Eq.\ref{computingP}).
	
In particular, when $\alpha = 2$, 
	the simple general form~(\ref{cooperative_optimization_general3})  in a continuous-time version is
\begin{equation}
-\hbar \frac{\partial \psi_i (x_i, t)}{\partial t} = \frac{1}{Z_i (t)} e_i(x_i) \psi_i (x_i, t) \ . 
\label{cooperative_optimization_general3.2}
\end{equation}
where
\[e_i(x_i)=\sum_{\sim x_i} \left( E_i (x) \prod_{j \not= i} |\psi_j(x_j, t)|^2  \right) \ . \]

Following the notation from physics, denote $\psi_i (x_i, t)$ as a vector $ \mid \psi_i (t) \rangle$.
Let $H_i$ be a diagonal matrix with diagonal elements as $e_i(x_i)$.
Then the equation~(\ref{cooperative_optimization_general3.2}) becomes
\begin{equation}
-\hbar \frac{d}{d t} \mid \psi_i (t) \rangle = \frac{1}{Z_i (t)} H_i \mid \psi_i (t) \rangle \ . 
\label{cooperative_optimization_general3.3}
\end{equation}
The above equation can be further generalized with a hermitian matrix $H_i$.

The expected return function $\psi_i (x_i, t)$ is also called a wavefunction in physics.
It is important to note that the equation~(\ref{cooperative_optimization_general3.3}) 
	is the dual equation of the Schr\"{o}dinger equation,
%\[ i \hbar \frac{d}{d t} \mid \psi_i (t) \rangle = H_i \mid \psi_i (t) \rangle \ , \] 
where $-1$ is replaced by the imaginary unit $i$ and the normalization factor $Z_i(t)$ is not required
	since the equation is unitary.
When the dynamic equation~(\ref{cooperative_optimization_general3.3}) 
	reaches a stationary point (equilibrium),
	the equation becomes the time-independent Schr\"{o}dinger equation:
\[ \lambda_i \mid \psi_i (x_i, t) \rangle = H_i \mid \psi_i (x_i, t) \rangle \ , \]
where $\lambda_i$ can only be any one of the eigenvalues of $H_i$.

%%%%%%%%%%%%%%%%%%%%%%%%%%%%%%%%%%%%%%%%%%%%%%%%%%%%%%%%%%%%%%%%%%%%%%%%%%%%%%%%%%%%
\begin{theorem}
When the parameter $\alpha = 2$, 
	the global optimization algorithm~(\ref{cooperative_optimization_general3}) in a continuous-time version
	becomes a dual equation of the Schr\"{o}dinger equation in quantum mechanics.
It falls back to the time-independent Schr\"{o}dinger equation whenever an equilibrium point is reached. 
\end{theorem}
%%%%%%%%%%%%%%%%%%%%%%%%%%%%%%%%%%%%%%%%%%%%%%%%%%%%%%%%%%%%%%%%%%%%%%%%%%%%%%%%%%%%

%The author is convinced that the algorithm is general 
%	and can be applied to solve optimization problems from different fields,
%	such as game theory described below.
    
%%%%%%%%%%%%%%%%%%%%%%%%%%%%%%%%%%%%%%%%%%%%%%%%%%%%%%%%%%%%
\section{From quantum mechanics to game theory}
%%%%%%%%%%%%%%%%%%%%%%%%%%%%%%%%%%%%%%%%%%%%%%%%%%%%%%%%%%%%

Let $u_i (x) = e^{-E_i(x)/\hbar}$ be the utility function for the player $i$ in a game of $n$ players.
In this case, the player $i$ tries to maximize his utility function $u_i(x)$ instead of 
	minimizing his objective function $E_i(x)$.
Both tasks are fully equivalent to each other.
In this case, (\ref{cooperative_optimization_general3}) becomes as
\begin{equation}
\Psi_i (x_i, t) = \sum_{\sim x_i} \left( u_i(x) \prod_{j \not= i} p_j(x_j, t-1) \right) \ , 
\label{cooperative_optimization_general4}
\end{equation}
\begin{equation}
\mbox{where }p_i(x_i, t) = \left(\Psi_i(x_i,t)\right)^{\alpha} / \sum_{x_i} \left(\Psi_i(x_i,t)\right)^{\alpha} \ . 
\label{mapping}
\end{equation}

In (\ref{mapping}), 
	when $\alpha \rightarrow \infty$, the best action $x_i$ for the player $i$ has a non-zero probability while others have probability zero.
That is, the player only accepts the best action, the one with the highest payoff $u_i(x_i, p_{-i})$.
In this case, the player is completely selfish.

If the value of $\alpha$ is reduced from the above extreme case, 
	the player $i$ starts to accept sub-optimal actions by assigning non-zero probability to them.
The degree of the acceptance increases with further decrease of $\alpha$.
At another extreme case, when $\alpha \rightarrow 0$,
	each action is assigned with the same probability 
	and the player has no preference on any one of the actions.
All of the actions are treated equally and they are sampled uniformly.
In this case, the player is completely selfishless.

In summary, the parameter $\alpha$ is kind of describing the selfishness level of player $i$. 
It covers the spectrum ranging from complete selfishness ($\alpha \rightarrow \infty$) to complete selfishlessness ($\alpha = 0$).

%%%%%%%%%%%%%%%%%%%%%%%%%%%%%%%%%%%%%%%%%%%%%%%%%%%%%%%%%%%%%%%%%%%%%%%%%%%%%%%%%%%%
\begin{theorem}
Based on Brouwer fixed point theorem, an equilibrium point always exists 
	for the set of equations~(\ref{cooperative_optimization_general4}) given any value for the selfishness level $\alpha$. 
It is still true even if each player $i$ in the game has his own selfishness level $\alpha_i$, possibly different from the rest.
\end{theorem}
%%%%%%%%%%%%%%%%%%%%%%%%%%%%%%%%%%%%%%%%%%%%%%%%%%%%%%%%%%%%%%%%%%%%%%%%%%%%%%%%%%%%
	
From Eq.~(\ref{mapping}), it is straightforward to prove that the point is also an $\epsilon$-approximate Nash equilibrium,
	where $\epsilon$ is inversely proportional to $\alpha$.
An $\epsilon$-approximate Nash equilibrium is a strategy profile such that no other strategy can improve
	the payoff by more than the value $\epsilon$.
In particular, a Nash equilibrium~\cite{nash50} can be viewed as an $0$-approximate one.

%%%%%%%%%%%%%%%%%%%%%%%%%%%%%%%%%%%%%%%%%%%%%%%%%%%%%%%%%%%%%%%%%%%%%%%%%%%%%%%%%%%%
\begin{theorem}
When the selfishness level $\alpha$ is sufficiently large, i.e., 
	$\alpha \rightarrow \infty$, 
	any equilibrium point for the set of iterative equations~(\ref{cooperative_optimization_general4})
	can be arbitrarily close to a Nash equilibrium and {\it vice versa}.
\end{theorem}
%%%%%%%%%%%%%%%%%%%%%%%%%%%%%%%%%%%%%%%%%%%%%%%%%%%%%%%%%%%%%%%%%%%%%%%%%%%%%%%%%%%%

Some experimental results are given in the following section to 
	demonstrate the improvement on the average individual payoff and the stability of game playing by reducing the selfishness level.

%------------------------------------------------------
\section{Experimental Results}
%------------------------------------------------------

An example payoff matrix of the prisoner's dilemma is given as follows:

\begin{center}
\begin{tabular}{|c|c|c|}
 \multicolumn{1}{c}{}& \multicolumn{1}{c}{Cooperate} &  \multicolumn{1}{c}{Defect}\\
\cline{2-3}
\multicolumn{1}{r|}{Cooperate} &  ~3,3~ & ~1,4~\\
\cline{2-3}
\multicolumn{1}{r|}{Defect} & ~4,1~ & {\bf 2,2} \\ 
\cline{2-3}
\end{tabular}
\end{center}

\begin{figure}
\centering
\center{\epsfxsize 8cm \epsffile{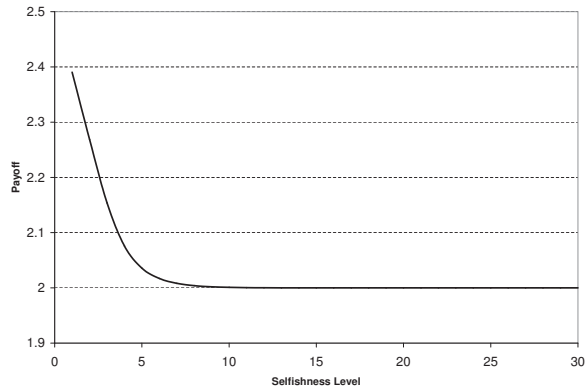}}
\caption{Payoffs of prisoner's dilemma under different selfishness levels.}
\label{fig-1}
\end{figure}

Payoffs of prisoner's dilemma under different selfishness levels are shown in Fig.~\ref{fig-1}.
When the selfishness level $\alpha$ reduces to one ($\alpha = 1$), 
	the payoff for each player has an $20\%$ improvement over the one at the Nash equilibrium ($\alpha = \infty$).

A 2-player game used in other game theory literatures has the following payoff matrix:
	
\[
\left(
\begin{array}{rrrrr}
 2,~3~ & ~-1,~4~ & ~2,~4~ & ~5,~2~ & ~1,-1 \\
 2,~2~ & ~3,~0~ & ~4,~1~ & ~-2,~4~ & ~1,~3 \\
 4,~6~ & ~7,~2~ & ~2,-2~ & ~4,~9~ & ~2,~1 \\
 9,~0~ & ~-2,~6~ & ~6,~3~ & ~7,~0~ & ~0,~5 \\
 3,~2~ & ~6,~1~ & ~2,~5~ & ~5,~3~ & ~1,~0 \\
\end{array}
\right)
\]

\begin{figure}
\centering
\center{\epsfxsize 8.0cm \epsffile{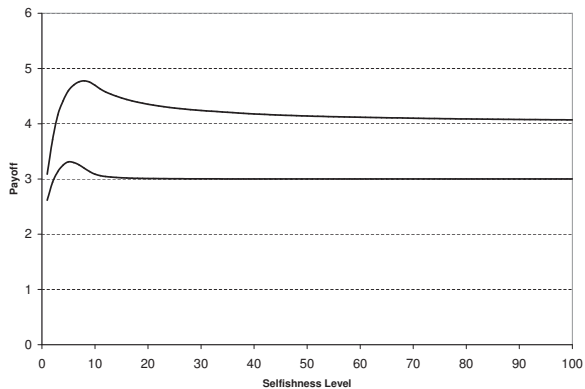}}
\caption{Payoffs of two players with 5 actions under different selfishness levels.}
\label{fig-2}
\end{figure}

This game has only one mixed Nash equilibrium.
Payoffs of the game under different selfishness levels $\alpha$ are shown in Fig.~\ref{fig-2}.
At $\alpha = 7$, the payoff of the row player has an $19\%$ improvement over the one at the Nash equilibrium
	and an $8.4\%$ improvement for the column player at the same time.

Computer-generated societies 
	are also used to test the impact of the selfishness level $\alpha$ on the payoffs of players and the stability of the societies.
In each computed generated society, each individual has a number of neighbors
	and his payoff function is defined by the summation of the pairwise joint actions of himself and his neighbors as follows
\begin{equation}
u_i (x) = \sum_{j \in {\cal N}(i)} f_{ij} (x_i, x_j), \quad \mbox{for $i=1,2,\ldots, n$} \ , 
\label{binary_payoff_function}
\end{equation}
where ${\cal N}(i)$ is the set of the individual $i$'s neighbors.
Each function value $f_{ij}(x_i, x_j)$ is uniformly sampled from the interval $[-0.6, 0.4]$.
The neighbors of each individual are randomly picked from the entire population.
The overall payoff of the society is defined as $u_1(x) + u_2(x) + \cdots + u_n(x)$.

\begin{figure}
\centering
\center{\epsfxsize 8cm \epsffile{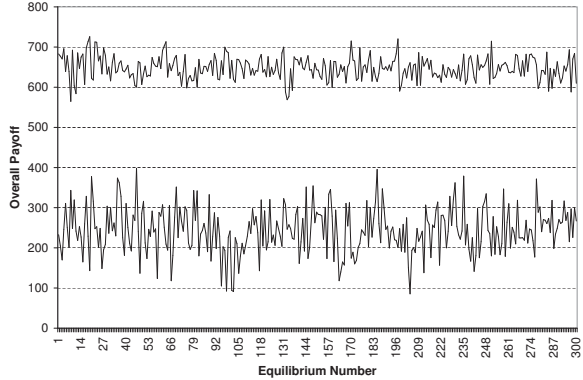}}
\caption{For a computer generated society of $1,001$ individuals, 
	when the individuals are completely selfish ($\alpha = \infty$),
	300 equilibria are found with their overall payoffs shown as the bottom connected lines.
When they are less selfish ($\alpha=30$),
	300 equilibria are also found with their overall payoffs shown as the top connected lines.
When $\alpha=\infty$, the average value of the overall payoffs is $243.62$ and the variance is $3,335$.
When $\alpha=30$, the corresponding values are $647.43$ and $818$.
}
\label{fig_8}
\end{figure}

In the first experiment,
	a society of $1,001$ individuals is generated where each one has $10$ actions and $50$ neighbors on average.
$300$ Nash equilbria ($\alpha=\infty$) are discovered together with $300$ ones under the selfishness level $\alpha=30$.
Fig.~\ref{fig_8} shows the overall payoffs of the first 300 ones versus the second 300 ones.
From the figure we can see that, reducing the selfishness level can lead to remarkable improvements both in the overall payoff and the stability 
	(the stability here is defined as the fluctuation of the overall payoffs of the equilibrium points).
At the same time, the quality of the overall payoffs has reached a new high level, 
	where the worst of the $300$ equilibria under the selfishness level $\alpha=30$ is still better than the best of the $300$ Nash equilbria
	in terms of the overall payoff.

\begin{figure}
\centering
\center{\epsfxsize 8cm \epsffile{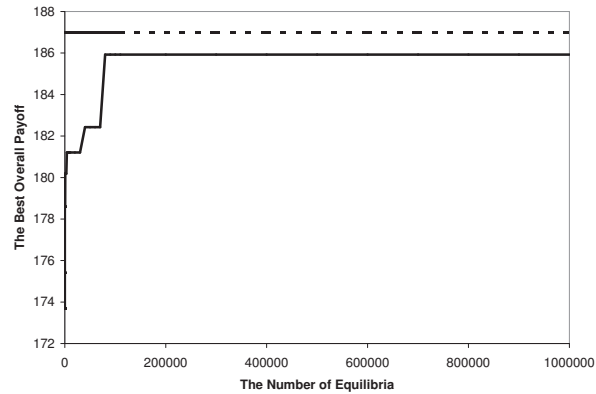}}
\caption{After exploring one million equilibria by a society of $121$ completely selfish individuals (the solid line), 
		the best one in terms of the overall payoff still couldn't match the single trial (averaged) by the same society when all the individuals are less selfish (the dotted line).}
\label{fig-9}
\end{figure}

A less selfish society can be more efficient 
	than a completely selfish society
	in terms of finding an equilibrium with a good overall payoff.
To compare the efficiency, 
	a society of a population of $121$ individuals with $50$ actions for each and $6$ neighbors on average is generated.
When the individuals in the society are less selfish ($\alpha = 20$),
	the average overall payoff of the 300 equilibria found by the society is $187$.
When they are completely selfish,
	after exploring one million of equilibria, 
	the best overall payoff is $185.9$, 
	still less than the former one (see Fig.~\ref{fig-9}).
%This result says that the average payoff of the less selfish society in an equilibrium under the selfishness level $\alpha$ 
%	is better than the best payoff out of those of one million Nash equilbria explored by the completely selfish society.
The less selfish society spent seconds on average to find an equilibria
	while the completely selfish society took almost a whole day to find the one million equilibria 
	using a laptop with a AMD Turion\texttrademark X2 Dual-Core Mobile Processor and 3GB RAM.
The former is much more efficient than the latter.

%---------------------------------------------------
\section{Conclusion}
%---------------------------------------------------
Accepting sub-optimal actions is a general decision-optimization principle for cooperation. 
It defines a global optimization approach to understand quantum mechanics.
It also offers a strategy for improving social stability and individual payoffs 
	over the classic profit-maximization principle.

The fundamental principle of rational decision making in classic game theory is to maximize the payoff by each player in a game. 
The logical justification of this principle seems obvious
	which shapes the definition of Nash equilibrium more than 50 years ago. 
However, the study presented in this paper
	shows that the optimality of this principle is conditional.
The optimal decision of each player in the classic sense 
	may not lead to a good payoff for the player.
Defying the conventional wisdom, compromising it by accepting sub-optimal actions can improve both the overall payoff,
	equivalently the average individual payoff,
	and the stability of game playing.
This study suggests that, for the benefit of everyone in a society (or a financial market), 
	the pursuit of maximal payoff by each individual should be controlled at some level
	either by voluntary good citizenship or by imposed regulations.

\end{document}